\documentclass[letterpaper, 10 pt, conference]{ieeeconf}  

\IEEEoverridecommandlockouts                              
\usepackage{amsmath, amssymb}
\usepackage{graphicx}
\usepackage{subfigure}

\usepackage{color}
\usepackage{times}
\usepackage{subfig}
\usepackage{cite}
\usepackage{stfloats}
\usepackage{xcolor, soul}
\sethlcolor{green}
\usepackage{mathtools}
\usepackage{balance}

\newtheorem{remark}{Remark}

\newtheorem{theorem}{Theorem}

\newtheorem{lemma}{Lemma}

\newtheorem{corollary}{Corollary}

\title{\LARGE \bf
Dynamic Modeling and Stability Analysis of Balancing in Riderless Electric Scooters
}

\author{{{Yun-Hao}} Lin, Alireza Jafari, and {{Yen-Chen}} Liu
\thanks{This work was supported in part by the National Science and Technology Council (NSTC), Taiwan, under Grant NSTC~112-2636-E-006-001 and NSTC~112-2628-E-006-014-MY3.}
\thanks{Y. -H. Lin, A. Jafari, and  Y.-C. Liu are with the Mechanical Engineering Department at National Cheng Kung University, Tainan, Taiwan.
        {\tt\small Email: N16110297@gs.ncku.edu.tw, alirezajafari110@gmail.com, yliu@mail.ncku.edu.tw}}%
}

\begin{document}

\maketitle
\thispagestyle{empty}
\pagestyle{empty}

\begin{abstract}
Today, electric scooter is a trendy personal mobility vehicle. The rising demand and opportunities attract ride-share services. A common problem of such services is abandoned e-scooters. An autonomous e-scooter capable of moving to the charging station is a solution. This paper focuses on maintaining balance for these riderless e-scooters.
The paper presents a nonlinear model for an e-scooter moving with simultaneously varying speed and steering. A PD and a feedback-linearized PD controller stabilize the model. The stability analysis shows that the controllers are ultimately bounded even with parameter uncertainties and measurement inaccuracy.
Simulations on a realistic e-scooter with a general demanding path to follow verify the ultimate boundedness of the controllers. 
In addition, the feedback-linearized PD controller outperforms the PD controller because it has narrower ultimate bounds. Future work focuses on experiments using a self-balancing mechanism installed on an e-scooter.

\end{abstract}

\section{INTRODUCTION}
In recent years, e-scooters have gained significant popularity due to their energy efficiency, low carbon footprint and convenience for the public.
The rising demand, coupled with rapid advancements in computing, creates an opportunity for the development of autonomous e-scooters.
For example, ride-sharing services benefit from autonomous e-scooters.
Abandoned e-scooters are often left for recycling by maintenance personnel, incurring significant human resource costs and planning issues~\cite{Turan2023}. 
Autonomous e-scooters are easier to recycle since they can autonomously navigate to charging stations~\cite{Ref_Wenzelburger2020}.
\textcolor{black}{Kondor et al. showed that the self-repositioning e-scooters' utilization is up to ten times higher, and they reduce the fleet size by an order of ten~\cite{Kondor2022}. In addition, self-balancing e-scooters and bicycles assist kids in learning how to ride~\cite{Ref_Cristiana2023} and elderly or people with disabilities in daily commutes.
Therefore, self-balancing e-scooters contribute to both public conveniences and ride-share economics.} 
Despite the incentives, the research on autonomous two-wheelers is sparse, and for the case of e-scooters, it is even sparser. 
Next, we review the related work to autonomous two-wheelers, focusing on the balancing problem.
\par
A dynamic model is essential in the balancing external torque design.
E-scooters' dynamics are similar to bicycles and motorcycles because they all have two wheels with distinct axes of rotation, and a handlebar controls the direction. 
Nonetheless, some differences require developing new models to improve their design and contribute to riders' and pedestrians' safety.
For example, Asperti et al. modeled the mechanical impedance of the rider to address vertical dynamics of e-scooters~\cite{Ref_Asperti2022}.
In addition, Garc{\'i}a-Vallejo et al. replaced a standard bicycle parameter set with an e-scooter’s set and reported that the e-scooters are fundamentally unstable and stable freehand ride is impossible~\cite{Ref_Garcia2020}.  
Although studies on bicycles have been around for decades, their dynamics are still an open issue for researchers.
For example, for decades, researchers believed that a riderless bicycle steers towards the fall to recover from it because of the gyroscopic precession and the negative trailing.
However, Kooijman et al. showed that their effects are insignificant, and the main contributors are the front mass location and the steering pole roll angle~\cite{Kooijman2011}.\par
Astrom et al. laid the foundation for recent bicycle studies~\cite{Ref_Astrom2005}. 
They started with a simple second-order model and step-by-step improved it by adding the front fork, modeling the rider's action by proportional feedback, gyroscopic effects of the front wheel, and rear-wheel steering.
Later, Xiong et al. used a reduced Whipple model to stabilize a bicycle by controlling the steering angle and the rear wheel speed~\cite{Ref_Xiong2021}.
Their controller is a function of the roll angle, changing the trajectory to keep the balance by design.
\textcolor{black}{Getz modeled a bicycle by the generalized Lagrange method and used torques applied on the steering pole and rear wheel to track a trajectory and maintain balance~\cite{Ref_Getz1995}. 
He simplifies the model by assuming certain limits on the roll angle, the steering angle, and their rates.
Chen et al. use the generalized Lagrange method and derive the bicycle dynamics considering the holonomic and nonholonomic constraints~\cite{Ref_Chen2006}. They develop a PID and a fuzzy controller to maintain the stability of a bicycle moving on a straight line or a circular track, i.e., constant steering.
Moreover, Zhang et al. model a riderless bicycle using multiple rigid bodies.
They linearize the model and analyze the factors influencing the bicycle's stability~\cite{Ref_Zhang2021}.
}
\par
The riders maintain the balance by changing their speed, steering in the proper direction, and creating an external torque using their body~\cite{Ref_Jafari2024}.
Similarly, self-balancing may utilize the same strategies, i.e., triggering a lean by steering and speed control or employing an external torque. 
However, changing the speed and steering interferes with the desired trajectory~\cite{Ref_Yetkin2014}.
Regarding maintaining balance by steering and speed control, Cui et al. divide autonomous bicycle control into two interconnected subsystems: tracking control and balance control~\cite{Ref_Cui2021}.
The balancing subsystem uses the steering angular velocity as the controller action. 
They prove the asymptotic stability of the coupled system using the small-gain theorem and perform simulations to validate its efficacy.
Wang et al. design a bikebot with a gyro-balancer~\cite{Ref_WANG2015, Ref_Wang2017, Ref_WANG2019}.
They focus on balancing the bikebot using only steering and speed control. 
However, using steering and speed control to maintain the balance sacrifices trajectory tracking. 
Thus, they add the gyro-balancer to provide assistive torque and relax the steering and speed actions in self-balancing.
\par
An alternative solution is using an external torque to keep the balance.
He et al. focus on autonomous bicycles and propose a Lyapunov-based tracking controller to follow the desired trajectory and a back-stepping-based controller to maintain the balance~\cite{Ref_He2022}.
The back-stepping controller applies external torque to the coupled dynamics of the bicycle and the tracking controller using a pendulum. 
Seekhao et al. use a pendulum and steering to balance a bicycle robot; they assume constant speed~\cite{Ref_Seekhao2020}.
They derived the nonlinear dynamics of their coupled system and then linearized it for stability analysis.
Their design balance a bicycle robot moving in a straight line.
Moreover, Wang et al. apply the gain scheduling on a bicycle using a momentum wheel~\cite{Ref_Wang2020}. 
They assumed a slight roll angle to linearize the trigonometric terms.
Soloperto et al. simulate an e-scooter maintaining balance with an external torque provided by a momentum wheel~\cite{Ref_Soloperto2021}.
In addition, they propose an algorithm to follow a trajectory and avoid obstacles using depth cameras.
\par
\textcolor{black}{Overall, autonomous e-scooters follow a trajectory~\cite{Ref_CARMONA2022} and maintain balance simultaneously.
Some prior research~\cite{Ref_Xiong2021, Ref_Getz1995, Ref_Cui2021} employed steering and speed control to maintain balance, which often sacrifices trajectory tracking and may require manouvers that are not feasible in real-world environments due to environmental constraints.
Others used external torque to maintain the balance~\cite{Ref_Chen2006,Ref_Zhang2021,Ref_Wang2017,Ref_He2022,Ref_Seekhao2020,Ref_Wang2020}.
However, they often linearized or simplified the model by assuming small roll angles, constant speed, slight steering angle, and negligible steering rate.
This paper focuses on self-balancing and assumes a higher level path-planning algorithm, e.g., the social force model in~\cite{T-ITS-I, SIMPAT2023}, determines the desired trajectory by dictating the speed and the steering angle.
Since we don't design the tracking sub-system, we use an external torque to maintain an e-scooter's balance with varying speed and steering angle inputs from the path planner.}
\par
The paper's contribution is to derive a novel nonlinear dynamic model, apply a Proportional-Derivative (PD) controller and a feedback-linearized PD, and stability analysis of the controllers.
Moreover, we perform simulations to verify the stability. 
The simplifying assumptions are helpful and can sufficiently explain e-scooter dynamics when the maneuvers are not demanding.
However, they affect the model's reliability during sudden or significant steering, high speeds, unexpected brakes, or harsh accelerations.
Therefore, we do not linearize the dynamics or assume any limitation on steering or speed except that they are continuous and differentiable.
\par
The remainder of the paper is structured as follows:
Section~\ref{sec:Dyn} presents the dynamics,
Section~\ref{sec:Cont} describes the controllers and their stability analysis,
Section~\ref{sec:Sim} discusses the simulations,
and Section~\ref{sec:Conc} summarizes the results and suggests future research direction.
\section{Dynamic Model}\label{sec:Dyn}
This section presents the dynamic model derived for bicycle-like two-wheelers, for example, the e-scooters.
First, we focus on the problem statement and clarify the objective.
In addition, a discussion on the system's holonomy explains why the standard form of Lagrange formulation suffices for our problem rather than the generalized form.
Then, the Lagrangian is used to derive the nonlinear dynamics.
\begin{figure}
\centering
\includegraphics[width=0.48\textwidth]{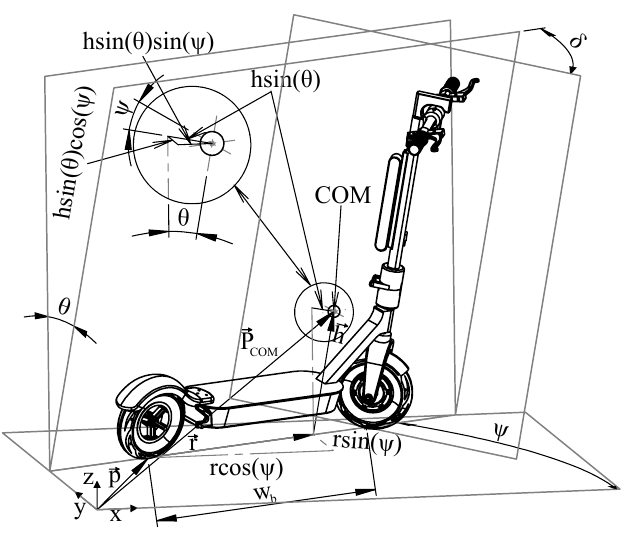}
\caption{E-scooter geometry in 3D space. The steering pole is normal to the e-scooter frame.}
\label{fig:Model:Geometry}
\end{figure}
\subsection{Preliminaries}
\subsubsection{Problem statement}
This research focuses on the e-scooter's balance when a higher-level controller or a path-planning algorithm determines the trajectory and the speed of a riderless e-scooter using steering angle $\delta$ and speed $v$ commands. In this paper, the speed is a scalar measured at the rear wheel, while the velocity is a vector measured at the Center of Mass (COM).
We do not make assumptions regarding constant speed or steering angles.
This study models the e-scooter's roll dynamics while the inputs, i.e., the speed and the steering angle, vary simultaneously.
\par
Fig.~\ref{fig:Model:Geometry} is a typical e-scooter performing a general maneuver with varying $\delta$, $v$, and the roll angle $\theta$.
In this figure, ${\dot \psi}$ and ${\Ddot \psi}$ are the rotational rates around the z-axis and $\delta$ and $\dot \delta$ are the steering angle and its rate.
Using ${\dot \psi}$ and ${\Ddot \psi}$ instead of $\delta$ and $\dot \delta$ simplifies the formulation.
Therefore, we apply the change of variables to the input.
Since the rear wheel contact point is the e-scooter's instantaneous center of rotation around the z-axis, 
\begin{align}
        &{\dot \psi}=\frac{v}{w_b}\tan{\delta}\label{eq:Map1},~\text{and}\\
        &{\ddot \psi}=\frac{v}{w_b}{\dot \delta} (1+\tan^2{\delta})+\frac{\dot v}{w_b}\tan{\delta},
            \label{eq:Map2}
\end{align}
where $w_b$ is the distance between the rear and front wheels' contact points or the wheelbase; the over-dots are the time derivatives.
Consequently, the inputs $v$ and $\delta$ and their first time derivatives uniquely determine ${\dot \psi}$ and ${\ddot \psi}$.
\par
\subsubsection{Discussion on holonomy}
Mathematically, for a set of generalized coordinates $q_1,q_2,\ldots,q_n$ describing a system's motion, the system is holonomic if the constraints are in the form $f(q_1,q_2,\ldots,q_n)=0$, where $f$ is a function of the coordinates only.
In contrast, a non-holonomic system has constraints that cannot be fully expressed using only the coordinates but require the consideration of their time derivatives~\cite{Ref_BLOCH1996}.
For example, two-wheeler kinematic constraints are in the form of $f(q_1,q_2,\ldots,q_n){\dot q}=0$, meaning that $q_i$s can be assigned arbitrarily, whereas the velocities are restricted~\cite{Ref_Limebeer2018}.\par
Because only the planar velocities affect the e-scooter's balance and not the positions, the self-balancing subset of equations only includes the velocities and does not rely on the coordinates.
Therefore, setting the planar velocities as the new generalized coordinates causes the constraints to be holonomic.
Thus, while the whole e-scooter kinematics is non-holonomic, the self-balancing subset of equations is holonomic. 
Consequently, the standard Lagrangian formulation applies to the self-balancing kinematics subset, unlike the whole kinematics, which requires the generalized Lagrangian and the Lagrange multipliers.
\par
\subsection{System modeling}
To set up the Lagrangian, we express the COM translational and rotational velocities as functions of the inputs and the roll angle.
Fig.~\ref{fig:Model:Geometry} is the e-scooter geometry in 3D space concerning coordinate system x-y-z; the steering pole is normal to the e-scooter frame.
Its angle with the x-axis in the x-y plane is $\psi$, and $\theta$ represents the e-scooter's roll angle.
Additionally, $r$ and $h$ are the horizontal and vertical distances of the COM from the rear wheel's contact point when the e-scooter is standing upright, i.e., $\theta=0$.
$\vec{P}_{COM}$, $\vec{p}$, $\vec{r}$, and $\vec{h}$ are the absolute position of the COM, the absolute position of the rear wheel's contact point, the relative position of COM to the rear contact point projected on the connecting line of rear and front wheels' contact points, and the relative position of the COM to the projection point, respectively.\par
Therefore, $\vec{P}_{COM}=\vec{p}+\vec{r}+\vec{h}$ or 
\begin{align}
        \vec{P}_{COM}=&\left({p}_{x}+r\cos{\psi}+h\sin{\theta}\sin{\psi}\right)\vec{i}\\
        &+\left({p}_{y}+r\sin{\psi}-h\sin{\theta}\cos{\psi}\right)\vec{j}\\
        &+{h\cos{\theta}}\vec{k},
            \label{eq:P_COM}
\end{align}
when presented by coordinate axes unit vectors $\vec i$, $\vec j$, and $\vec k$; subscripts x, y, and z are variable components along the corresponding axis.
\par
Therefore, the velocity components of the COM are 
\begin{align}
    &{V}_{x}={\dot p}_{x}-r{\dot \psi}\sin{\psi}+h{\dot \theta}\cos{\theta}\sin{\psi}+h{\dot \psi}\sin{\theta}\cos{\psi}\nonumber\\
    &{V}_{y}={\dot p}_{y}+r{\dot \psi}\cos{\psi}-h{\dot \theta}\cos{\theta}\cos{\psi}+h{\dot \psi}\sin{\theta}\sin{\psi}\nonumber\\
    &{V}_{z}={-h{\dot \theta}\sin{\theta}}.
    \label{eq:Pd_COM}
\end{align}
where ${\dot p}_{x}=v\cos{\psi}$ and ${\dot p}_{y}=v\sin{\psi}$.\par
The kinetic energy $T=\frac{1}{2}mV_c^2+\frac{1}{2}I_{\theta}{\dot \theta}^2+\frac{1}{2}I_{\psi}{\dot \psi}^2$,
and the potential energy $W=mgh\cos{\theta}$,
form the Lagrangian $L$ as
\begin{align}
     L&=T-W=\frac{1}{2}mV_c^2+\frac{1}{2}I_{\theta}{\dot \theta}^2+\frac{1}{2}I_{\psi}{\dot \psi}^2-mgh\cos{\theta},
\end{align}
where $V_c^2=V_x^2+V_y^2+V_z^2$ and $g$ is the gravitational constant.
$m$, $I_{\theta}$, and $I_{\psi}$ denote the e-scooter's mass, the roll and the yaw moments of inertia, respectively.\par
Since the $\delta$ and $v$, and subsequently, $\psi$, are inputs to the self-balancing subset, the generalized coordinate in the Lagrangian is $\theta$.
Hence, for the self-balancing subset,
\begin{align}
     &\frac{d}{dt}\left(\frac{\partial L}{\partial {\dot \theta}}\right)-\frac{\partial L}{\partial {\theta}}=\nonumber\\
     &(I_{\theta}+mh^2)\Ddot{\theta}-mhr\Ddot{\psi}\cos{\theta}-mgh\sin{\theta}\nonumber\\
     &-mh{\dot \psi}\left(v-h{\dot \psi}\sin{\theta}\right)\cos{\theta}=\tau_{\theta},
     \label{eq:Dynamics}
\end{align}
where $\tau_{\theta}$ is the external torque on the roll angle $\theta$. An external mechanism, for example a momentum wheel~\cite{Ref_Wang2020}, a gyroscope~\cite{Ref_Wang2017}, or a pendulum~\cite{Ref_He2022}, creates the external torque to maintain the balance. In~\eqref{eq:Dynamics},
$\dot \psi$ and $\ddot \psi$ are calculated using~\eqref{eq:Map1} and~\eqref{eq:Map2}.
By defining 
\begin{align}
     &M=I_{\theta}+mh^2,\nonumber\\
     &C=mhr\Ddot{\psi}+mh{\dot \psi}\left(v-h{\dot \psi}\sin{\theta}\right),\text{~and}\nonumber\\
     &G=mgh,\nonumber
\end{align}
\eqref{eq:Dynamics} results in
\begin{align}
     &M\Ddot{\theta}=\tau_{\theta}+C\cos{\theta}+G\sin{\theta}.
     \label{eq:DynPardef}
\end{align}
which we rewrite as
\begin{align}
     &M\Ddot{\theta}=\tau_{\theta}+U\sin{(\theta+\theta_0)},
     \label{eq:DynFinal}
\end{align}
where $U=\sqrt{C^2+G^2}$ and $\tan{\theta_0}=\frac{C}{G}$.\par
Overall,~\eqref{eq:DynFinal} is the roll dynamics obtained using the Lagrange method.
The next section introduces controllers to maintain the balance for~\eqref{eq:DynFinal}.\par
\section{Controller Design}\label{sec:Cont}
This section focuses on controller design and stability analysis for the obtained dynamics in the previous section.
Although a controller can follow a desired trajectory, we assume $\theta_d=0$ or the e-scooter's upright position is favorable for simplicity.
First, we apply a PD controller and prove that the states are ultimately bounded. 
Next, we employ the feedback linearization technique to improve the performance.
In an ideal situation, when the information is perfect, the former can only guarantee boundedness, whereas the latter ensures asymptotic stability.
With imperfect measurements and approximate modeling, the boundedness is still guaranteed for both controllers.
Nevertheless, the latter has smaller bounds on $\theta$, which is appealing.
In addition, the bounds are calculated for both cases. 
\par
\begin{figure}
\centering
\includegraphics[width=0.48\textwidth]{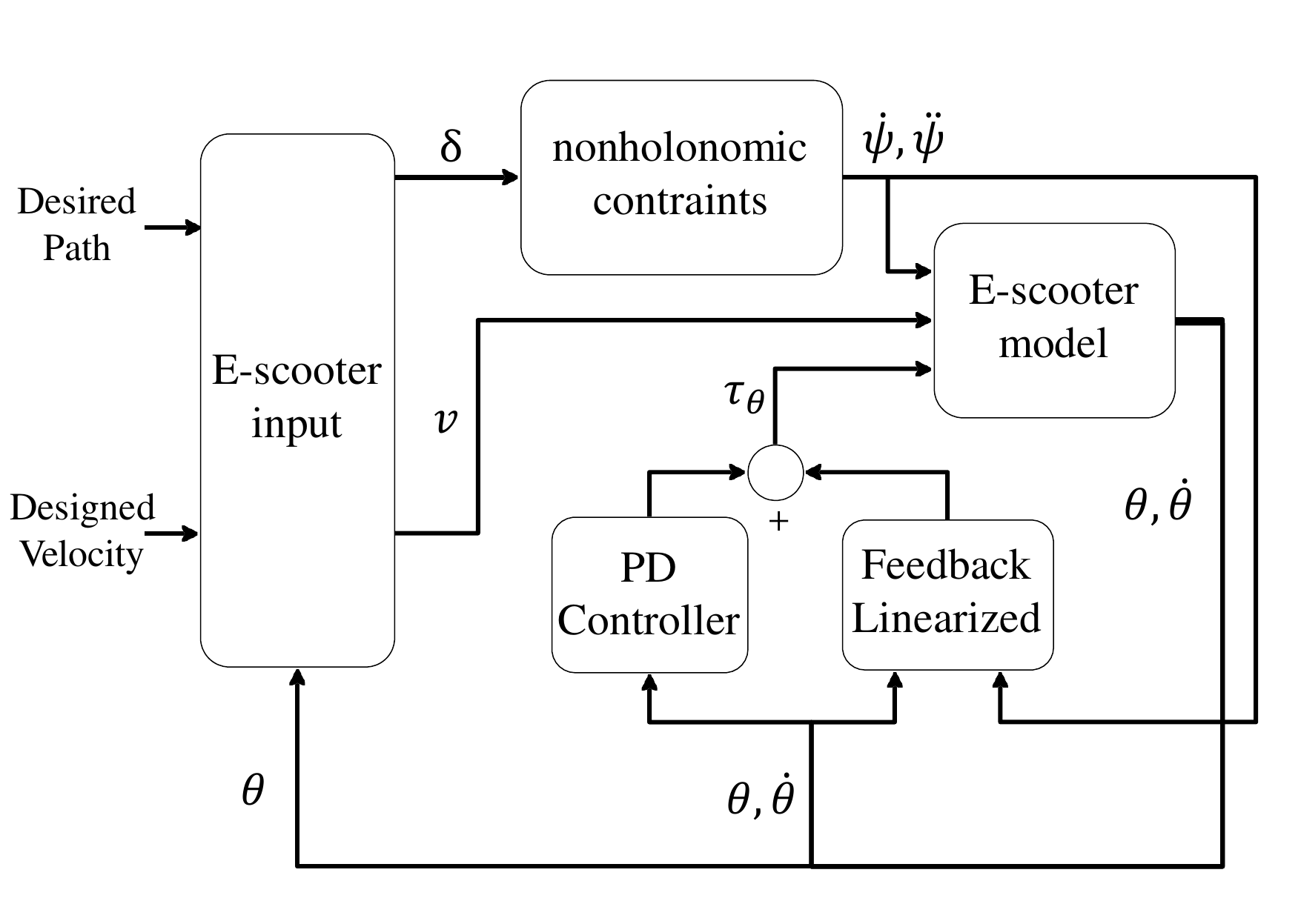}
\caption{Control block diagram of the self-balancing e-scooter}
\label{fig:Model:blockdiagram}
\end{figure}

Fig.~\ref{fig:Model:blockdiagram} is the system's block diagram.
The path planner converts the desired path and the desired speed to the desired steering angle $\delta$, rear wheel speed $v$, and their time derivatives $\dot \delta$ and $\dot v$.
The controller uses \eqref{eq:Map1} and \eqref{eq:Map2} to calculate $\dot \psi$ and $\ddot \psi$ and create the external torque $\tau_\theta$ by an external mechanism to keep the e-scooter in an upright position, i.e.,  $\theta_{d}=0$.
\subsection{PD Controller}\label{sec:PD}
In this section, we apply a PD controller to the obtained dynamics~\eqref{eq:DynFinal} and prove the boundedness of $\theta$ and $\dot \theta$.
In the first step, we present and prove Lemma~\ref{lemma1}.
\begin{lemma}\label{lemma1}
Consider the e-scooter dynamics~\eqref{eq:DynFinal}. The PD controller
\begin{align}
     &\tau_{\theta}=-K_d{\dot \theta}-K_p\theta,
     \label{eq:PD}
\end{align}
with positive $K_d$ and $K_p$ ensures that $\dot \theta$ is ultimately bounded by ${|{\dot \theta}|}_{max}={U}/{K_d}$.
\end{lemma}
\begin{proof}\label{prooflemma1}
To prove the boundedness, we introduce the Lyapunov function $V_1$,
\begin{align}
     &V_1(\theta, {\dot \theta})=\frac{1}{2}M{\dot \theta}^2+\frac{1}{2}K_p\theta^2.
     \label{eq:V1}
\end{align}
Taking the differentiation of $V_1$ with respect to time gives
\begin{align}
     &{\dot V}_1=M{\ddot \theta}{\dot \theta}+K_p{\dot \theta}\theta,
     \label{eq:V1d1}
\end{align}
which after substituting~\eqref{eq:DynFinal} and ~\eqref{eq:PD} and simplifying, results in
\begin{align}
     &{\dot V}_1=-K_d{\dot \theta}^2+U\sin{(\theta+\theta_0)}{\dot \theta}.
     \label{eq:V1d2}
\end{align}
If $|{\dot \theta}|>\frac{U}{K_d}$, then ${\dot V}_1<0$. 
We define 
\begin{align}
    B_{1,in}= \{(\theta,{\dot \theta})) \mid |{\dot \theta}|\leq\frac{U}{K_d}\}.
\end{align}
All the $V_1$s starting outside of $B_{1,in}$ enter the region within a finite time and remain inside. Since ${\dot V_1}$ is negative on the boundary, ${\dot \theta}$ is ultimately bounded and its upper bound is ${|{\dot \theta}|}_{max}={U}/{K_d}$.
\end{proof}
\par
\begin{remark}\label{remark1}
    According to Lemma~\ref{lemma1}, if $\dot \theta>\frac{U}{K_d}$, then $\theta$ and $\dot \theta$ are ultimately bounded. if $\dot \theta \leq\frac{U}{K_d}$, the ultimate boundedness of $\dot \theta$ is trivial, but the boundedness of $\theta$ is not guaranteed.
\end{remark}\par
Since the boundedness of $\dot \theta$ does not necessarily imply that $\theta$ is bounded, we present Theorem~\ref{theorem1} proving the boundedness of $\theta$ using Lemma~\ref{lemma1}.
\begin{theorem}\label{theorem1}
    Consider the e-scooter dynamics~\eqref{eq:DynFinal} and the PD controller~\eqref{eq:PD} with positive $K_d$ and $K_p$.
    Then, $\theta$ is ultimately bounded and it's bound is $U\left(\frac{K_d+\sqrt{\Delta}}{2K_dK_p}\right)$, where $\Delta=K_d^2+4K_pM>0$.
\end{theorem}
\begin{proof}\label{prooftheorem1}
    To prove the boundedness of $\theta$, we introduce a second Lyapunov function $V_2$ as
\begin{align}
     &V_2=\frac{1}{2}M({\dot \theta}+\lambda\theta)^2+\frac{1}{2}K\theta^2,
     \label{eq:V2}
\end{align}
where $\lambda>0$ and $K>0$ are constants to be assigned in the following steps. 
Differentiating $V_2$ with respect to time gives
\begin{align}
     &{\dot V}_2=M{\ddot \theta}({\dot \theta}+\lambda\theta)+M\lambda{\dot \theta}^2+M\lambda^2{\dot \theta}\theta+K{\dot \theta}\theta.
     \label{eq:V2d1}
\end{align}
By substituting~\eqref{eq:DynFinal} and~\eqref{eq:PD}, simplifying, and regrouping the terms, we get
\begin{align}
     {\dot V}_2=&\left((-K_d+M\lambda){\dot \theta}^2+U\sin{(\theta+\theta_0)}{\dot \theta}\right)\nonumber\\
     &+\left(-Kp\lambda\theta^2+\lambda U\sin{(\theta+\theta_0)}\theta\right)\nonumber\\
     &+\left((K+M\lambda^2-K_d\lambda-K_p){\dot \theta}\theta\right).
     \label{eq:V2d2}
\end{align}
Next, we assign 
\begin{align}
     K=-M\lambda^2+K_d\lambda+K_p,
     \label{eq:Kassign}
\end{align}
to eliminate the last term in~\eqref{eq:V2d2}. Positive $K$ entails
\begin{align}
     -M\lambda^2+K_d\lambda+K_p>0,
     \label{eq:Kassign2}
\end{align}
which requires 
\begin{align}
     &\Delta=K_d^2+4K_pM>0,  \label{eq:Conditions1}\text{and}\\
     &\frac{-K_d-\sqrt{\Delta}}{2}<\lambda<\frac{-K_d+\sqrt{\Delta}}{2}.
     \label{eq:Conditions2}
\end{align}
Since \eqref{eq:Conditions1} always holds and $\lambda>0$ satisfying~\eqref{eq:Conditions2} exists ($|K_d|<\sqrt{\Delta}$), we can assign $K>0$ by using~\eqref{eq:Kassign}. Thus, ${\dot V}_2$ simplifies to
\begin{align}
     {\dot V}_2=&\left((-K_d+M\lambda){\dot \theta}^2+U\sin{(\theta+\theta_0)}{\dot \theta}\right)\nonumber\\
     &+\left(-K_p\lambda\theta^2+\lambda U\sin{(\theta+\theta_0)}\theta\right).
     \label{eq:V2d3}
\end{align}
Since ${\dot \theta}$ is bounded, based on Lemma~\ref{lemma1}, the first term is bounded by $M\lambda\frac{U^2}{K_d^2}$, and therefore, $\dot V_2<0$ if $\theta>{|\theta|}_{max}$, 
\begin{align}
     {|\theta|}_{max}=U\left(\frac{K_d+\sqrt{\Delta}}{2K_dK_p}\right).
     \label{eq:V2d4}
\end{align}
We define 
\begin{align}
    B_{2,in}= \{(\theta,{\dot \theta}) \mid |{\theta}|\leq|\theta|_{max}\}.
\end{align}
Therefore, all the $V_2(\theta, {\dot \theta})$s starting outside of $B_{2,in}$ enter the region within a finite time and remain inside, since ${\dot V_2}$ is negative on the boundary.
Thus, ${\theta}$ is ultimately bounded and its upper bound is $U\left(\frac{K_d+\sqrt{\Delta}}{2K_dK_p}\right)$.
\end{proof}
\begin{remark}\label{remark2}
    Lyapunov function $V_1$ is a special case of Lyapunov function $V_2$ where $\lambda=0$.
\end{remark}
\par
Overall, for a PD controller, the bounds on $\theta$ and $\dot \theta$ depend on the inputs $\delta$ and $v$ and can be significant during demanding maneuvers.
Section~\ref{sec:FL+PD} suggests a feedback-linearized PD controller addressing the issue.
\par
\subsection{Feedback-linearized PD Controller}\label{sec:FL+PD}
During demanding maneuvers, the path-planner's inputs are notable, i.e., significant $v$, $\delta$, $\dot v$, and $\dot \delta$, $C$, and consequently $U$; $G$ is constant.
A large $U$ leads to undesirable values of $\theta$ and $\dot \theta$ and degrades the system's performance. 
To address the issue, we apply a feedback linearizer in addition to the PD controller.\par
We define $\tau_\theta$ as
\begin{align}
     &\tau_{\theta}=-K_d{\dot \theta}-K_p\theta-\hat{C}\cos{\theta}-\hat{G}\sin{\theta},
     \label{eq:PD+FL}
\end{align}
where $\hat{C}$ and $\hat{G}$ are estimations of $C$ and $G$ and subjected to parameter uncertainties and measurement inaccuracies.  
Therefore, using the estimations errors $\Tilde{G}=G-\hat{G}$ and $\Tilde{C}=C-\hat{C}$ and~\eqref{eq:DynPardef} and~\eqref{eq:PD+FL}, the new system is
\begin{align}
     &M{\ddot \theta}=-K_d{\dot \theta}-K_p\theta+\Tilde{C}\cos{\theta}+\Tilde{G}\sin{\theta},
     \label{eq:PD+FL2}
\end{align}
or
\begin{align}
     &M\Ddot{\theta}=-K_d{\dot \theta}-K_p\theta+\Tilde{U}\sin{(\theta+\Tilde{\theta}_0)},
     \label{eq:PD+FL3}
\end{align}
where $\Tilde{U}=\sqrt{\Tilde{C}^2+\Tilde{G}^2}$ and $\tan{\Tilde{\theta}_0}=\frac{\Tilde{C}}{\Tilde{G}}$. \par
Since~\eqref{eq:PD+FL3} has the same form as~\eqref{eq:DynFinal}, Section~\ref{sec:PD}'s discussions apply to~\eqref{eq:PD+FL}.
The difference is that the bounds are much smaller in practice because they depend on $\Tilde{U}$ instead of $U$.\par
\begin{corollary}
     With perfect estimations and known dynamics, i.e., $\Tilde{U}=0$, for the system described by dynamics~\eqref{eq:DynFinal} and the controller~\eqref{eq:PD+FL}, the bounds on $\dot \theta$ and $\theta$ are zero, and therefore, the system is asymptotically stable.
\end{corollary}
\par
Section~\ref{sec:Sim} simulates introduced controllers on the obtained dynamics to compare the performance.
\section{Simulations and discussions}\label{sec:Sim}

\subsection{Simulation Setup}\label{sec:sim:set}
\begin{figure}
\centering
$\begin{array}{c}
\subfigure[] {
\includegraphics[width=0.48\textwidth]{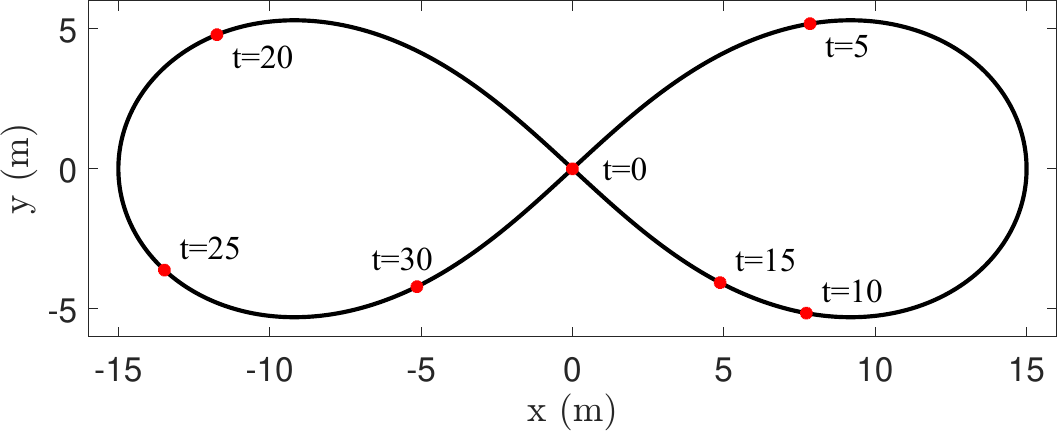}} \\
 \subfigure[] {
 \includegraphics[width=0.48\textwidth]{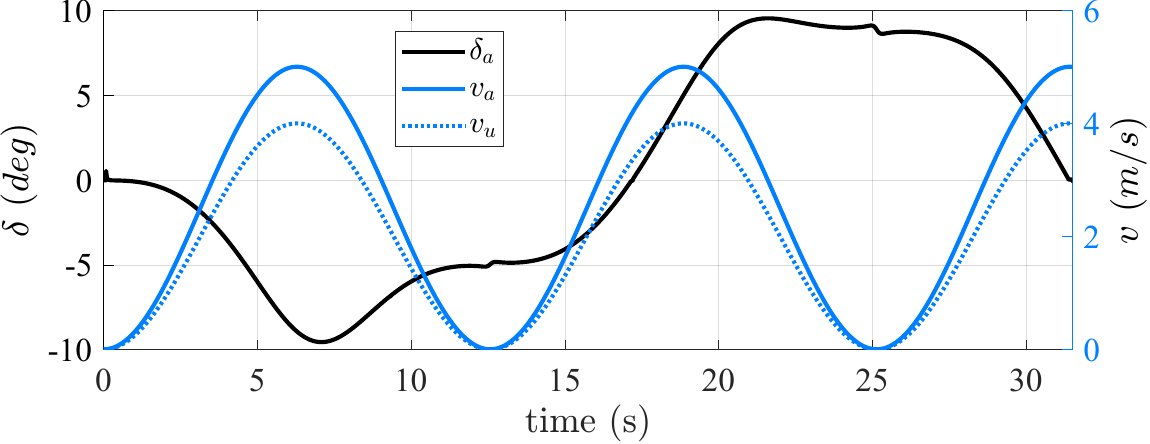}} 
\end{array}$
\caption{The simulation input: (a) desired path;
(b) the desired velocity and the steering angle created by the desired path and designed velocity.
}
\label{fig:Sim:In}
\end{figure}

Towards simulating with simultaneously varying steering angle and speed, we assume that Fig.~\ref{fig:Sim:In}(a), the Lemniscate of Bernoulli with $a=15$~\cite{Ref_Lockwood1961}, is the desired trajectory. The desired speed is $v = 2.5+2.5\sin{(\frac{1}{2}t+\frac{3}{2}\pi)}$ m/s; the path-planning algorithm sends the information at each instance.
Fig.~\ref{fig:Sim:In}(b) is the desired speed and corresponding steering angle realizing the desired trajectory.
Dozza et al. experimentally measured an e-scooter's braking ability during harsh brakes and reported 0.7$\pm$0.25 m/s$^2$~\cite{Ref_Dozza2023}.
Thus, the assumed desired speed in Fig.~\ref{fig:Sim:In}(b), with a maximum deceleration/acceleration of 1.125 m/s$^2$ and the speed range of 0 to 5 m/s, emulates the harsh braking and accelerating cases.
Moreover, the associated steering along the desired speed presents a demanding maneuver for the simulated e-scooter.
\par

\begin{figure}
\centering
$\begin{array}{c}
\subfigure[] {
\includegraphics[width=0.45\textwidth]{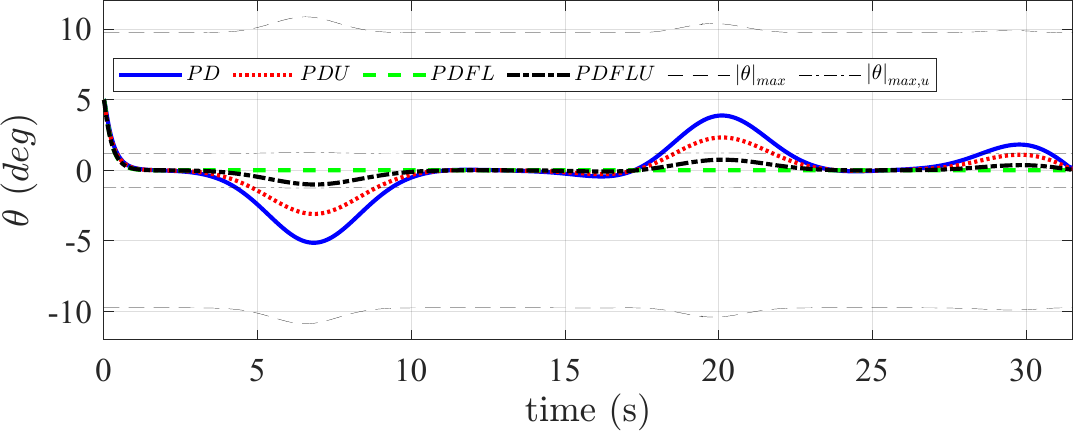}} \\
 \subfigure[] {
 \includegraphics[width=0.45\textwidth]{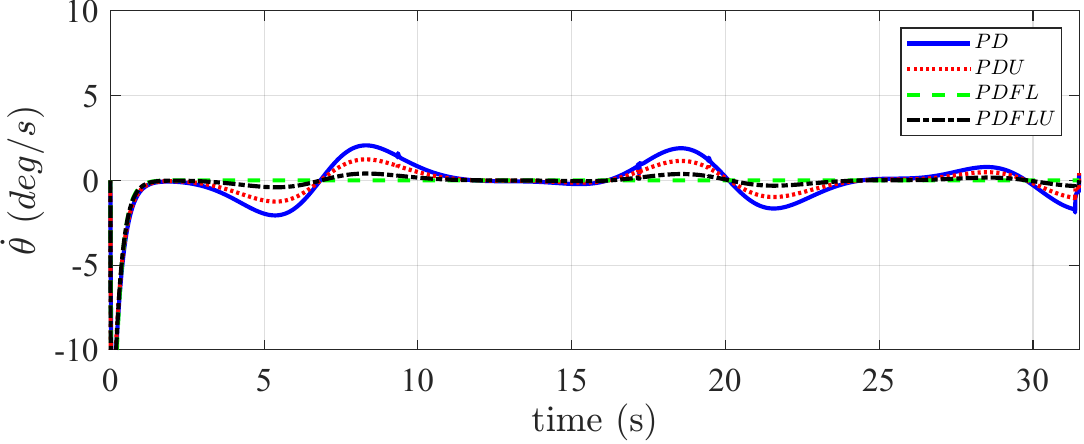}} \\
 \subfigure[]{
 \includegraphics[width=0.45\textwidth]{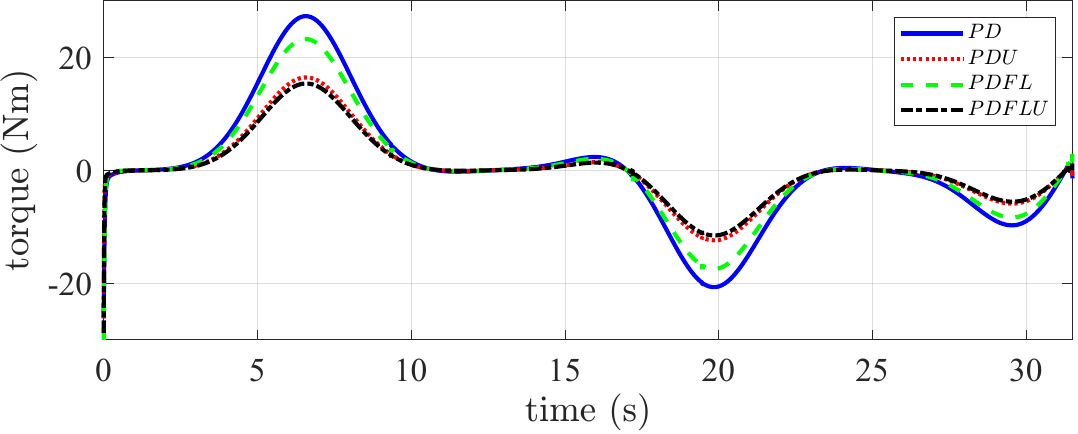}}
\end{array}$
\caption{The simulation results: (a) the roll angle $\theta$;
(b) the roll angle rate $\dot \theta$;
(c) the controller action $\tau_\theta$.
The legend and the abbreviations apply to all; 
PD controller with no uncertainties is labeled as "PD";
PD controller with uncertainties is labeled as "PDU";
PD Feedback Linearized controller with no uncertainties is labeled as "PDFL";
PD Feedback Linearized controller with uncertainties is labeled as "PDFLU"; $\left | \theta \right |_{max} $ represents the upper limit of the system's roll angle for PD; $\left | \theta \right |_{max,u} $ represents the upper limit of the system's roll angle for PDFLU.
}
\label{fig:Sim:Res}
\end{figure}

We use Segway ES4 e-scooter parameters to have a realistic simulation, presented in Table~\ref{Tab:Sim:Par}.
In addition, to simulate the effects of the measurement inaccuracies, we assume that $\delta$ and $v$ fed to the controller are different from the actual ones. 
Moreover, due to the uncertainties, $m$, $r$, and $h$ used in the controller differ from the actual ones.
The actual values and the approximate values are presented in Table~\ref{Tab:Sim:Par} with subscripts $a$ and $u$. 
We apply the actual values in the model and the approximate values in the controller.
\subsection{Simulation Results}\label{sec:sim:res}
This section presents simulation results for the PD controller and the feedback-linearized PD with and without uncertainties for a general case described in Section~\ref{sec:sim:set}. 
Fig.~\ref{fig:Sim:Res} shows the trajectories for $\theta$, Fig.~\ref{fig:Sim:Res}(a), $\dot \theta$, Fig.~\ref{fig:Sim:Res}(b), and the control actions $\tau_\theta$, Fig.~\ref{fig:Sim:Res}(c).
\par
Fig.~\ref{fig:Sim:Res}(a) is the roll angle trajectory for the controllers with and without uncertainties. It also shows the obtained bounds on $\theta$, i.e., \eqref{eq:V2d4} with $U$ for the PD without uncertainty and $\tilde{U}$ for feedback-linearized PD with uncertainty.
The trajectories are ultimately bounded and stay inside the obtained limits once they enter the region. However, the bounds for the PD controller are wider, allowing for significant fluctuations. Moreover, feedback-linearized PD converges faster and has much smaller bounds than PD. 
\textcolor{black}{According to Lemma~\ref{lemma1} and \eqref{eq:V2d4}, the bounds depend on $U$ and $\tilde{U}$ for the PD and the feedback-linearized PD, respectively.
Since $\tilde{U}\leq U$, the feedback-linearized PD bounds are smaller than PD bounds.
In addition,} when there is no uncertainty, the feedback linearized PD asymptotically converges to the origin.
\par
In addition, in Fig.~\ref{fig:Sim:Res}(b), $\dot \theta$ is ultimately bounded for both controllers and with and without uncertainties. In comparison, $\dot \theta$ is smaller for the feedback-linearized PD. 
Regarding the controller actions in Fig.~\ref{fig:Sim:Res}(c), the feedback-linearized PD achieves faster convergence and tighter bounds on the states with less controller action. Thus, the feedback-linearized PD performs better than the PD controller, even in the presence of uncertainties.
\par
\begin{remark}
    The peaks in the trajectories and the controller action happen when the desired speed and the steering are high simultaneously, e.g., $t\approx7$ in Fig.~\ref{fig:Sim:In}(b).
\end{remark}
\par
In conclusion, although previous research shows that the PD controller can guarantee the e-scooter balance with satisfactory performance, our simulations show that this differs for severe maneuvers and the nonlinear model. Although the PD controller is ultimately bounded, the bounds during demanding manouvers are large, and the performance is poor. However, the feedback linearized PD is still asymptotically stable and, even in the presence of uncertainty, outperforms the PD controller because of its narrower ultimate bounds.

\begin{center}
\begin{table}
\centering
\caption{Simulation parameters. The actual and approximate values have subscripts $a$ and $u$. We apply the actual values in the model and use the approximate values in the controller.}
\renewcommand{\arraystretch}{1.5}
\begin{tabular}{||p{0.11\textwidth}|p{0.11\textwidth}|p{0.19\textwidth}||}
\hline
 Description              &  Symbol~--~Unit       & Value \\ [0.5ex]
 \hline \hline
Initial roll          & \(\theta(0)\)~--~deg  &  10     \\ [0.5ex]
 \hline
Initial yaw           & \(\psi(0)\)~--~deg    &  0     \\ [0.5ex]
 \hline
Speed                       & $v_a$~--~m/s            &    $v = 2.5+2.5\sin{(\frac{1}{2}t+\frac{3}{2}\pi)}$  \\ [0.5ex]
 \hline
COM height   & $h_a$~--~m              &   0.34   \\ [0.5ex]
 \hline
COM distance  & $r_a$~--~m              &    0.63  \\ [0.5ex]
 \hline
Mass              & $m_a$~--~kg             &  14    \\ [0.5ex]
 \hline
Wheelbase             & $w_b$~--~m    &  0.84  \\ [0.5ex]
 \hline
Mom. of inertia  & $I_\theta$~--~kg.m$^2$    & 0.54 \\ [0.5ex]
 \hline
Steering angle              & \(\delta_a\)~--~deg    &  See Fig.~\ref{fig:Sim:In}(b)  \\ [0.5ex]
 \hline
Proportional gain            & $K_p$~--~N.m/rad    &  300  \\ [0.5ex]
\hline
Derivative gain             & $K_d$~--~N.m.s/rad    &  80  \\ [0.5ex]
 \hline
 \multicolumn{3}{||c||}{\centering Uncertain and inaccurate parameters in the simulations} \\ [0.5ex]
 \hline
Speed              & $v_u$~--~m/s             &  $0.8v_a$; see Fig.~\ref{fig:Sim:In}(b)   \\ [0.5ex]
 \hline
COM height  & $h_u$~--~m              &   0.27   \\ [0.5ex]
 \hline
COM distance & $r_u$~--~m              &    0.50  \\ [0.5ex]
 \hline
Mass              & $m_u$~--~kg             &  11.2    \\ [0.5ex]

 \hline
\end{tabular}
\label{Tab:Sim:Par}
\end{table}
\end{center}
\section{Conclusion and future works}\label{sec:Conc}
The paper presents a dynamic model for an e-scooter's self-balance when the steering angle and the speed change simultaneously.
We apply a PD and feedback-linearized PD controller to the model and prove their ultimate boundedness.
We also analyze the feedback-linearized PD controller stability in the presence of uncertainties.
Simulations verify the ultimate boundedness and compare the performance. The feedback-linearized PD has a higher convergence rate, and the states stay closer to the origin. In addition, it achieves higher performance with less controller effort.\par
Ongoing research focuses on implementing the controllers on a self-balancing e-scooter. The work includes adding a balancing mechanism on an e-scooter and updating the model accordingly. Future research direction is environment perception and path-planning for a self-balancing e-scooter cruising on a sidewalk with pedestrians.
Another promising direction is incorporating time-to-collision into the path planning algorithm to ensure pedestrian safety and comfort~\cite{IFAC2023,Ref_Jafari2024}.

	\bibliographystyle{IEEEtran}
	\bibliography{IEEEexample}

\end{document}